\documentclass[twocolumn]{aastex631}

\begin{document}

\title[A 30 day optical flare from AT2022zod]{AT2022zod: An Unusual Tidal Disruption Event in an Elliptical Galaxy at Redshift 0.11}

\author{Kristen C. Dage}

\affiliation{International Centre for Radio Astronomy Research—Curtin University, GPO Box U1987, Perth, WA 6845, Australia}
\author{Phelipe Darc}
\affiliation{Artificial Intelligence for Physics Laboratory (Lab-IA) and Centro Brasileiro de Pesquisas F\'isicas, Rua Xavier Sigaud, 150, Urca, Rio de Janeiro, Brazil}
\author{Thallis Pessi}
\affiliation{European Southern Observatory, Alonso de Córdova 3107, Vitacura, Casilla 19001, Santiago, Chile}
\author{Rupesh Durgesh}
\affiliation{Independent researcher, Germany}
\author{Shravya Shenoy}
\affiliation{ Centre for Astrophysics Research, University of Hertfordshire, Hatfield AL10 9AB, UK}
\author{Lilianne Nakazono} 
\affiliation{Observatorio Nacional / MCTI, Rua General Jos\'{e} Cristino 77, Rio de Janeiro, RJ, 20921-400, Brazil}
\affiliation{Departamento de Física Matem\'{a}tica, Instituto de F\'{i}sica, Universidade de S\~{a}o Paulo, R. do Mat\~{a}o 1371, 05508-090, S\~{a}o Paulo, SP, Brazil}
\author{Celine Boehm}
\affiliation{School of Physics, The University of Sydney and ARC Centre of Excellence for Dark Matter Particle Physics, NSW 2006, Australia}
\author{Ana L. Chies-Santos} 
\affiliation{Instituto de F\'isica, Universidade Federal do Rio Grande do Sul, Porto Alegre, RS 90040-060, Brazil}
\author{Rafael S. de Souza} 
\affiliation{ Centre for Astrophysics Research, University of Hertfordshire, Hatfield AL10 9AB, UK}
\affiliation{Instituto de F\'isica, Universidade Federal do Rio Grande do Sul, Porto Alegre, RS 90040-060, Brazil}
\affiliation{ Department of Physics \& Astronomy, University of North Carolina at Chapel Hill, NC 27599-3255, USA}
\author{Zs\'ofia V.  Kovács-Stermeczky}
\affiliation{ELTE E\"otv\"os Lor\'and University, Institute of Geography and Earth Sciences, Department of Astronomy, Budapest, Hungary}
\affiliation{HUN-REN Research Centre for Astronomy and Earth Sciences, Konkoly Observatory, MTA Centre of Excellence, Konkoly-Thege M.\'ut 15-17, H-1121, Budapest, Hungary}
\author{Arash Bahramian} 
\affiliation{International Centre for Radio Astronomy Research—Curtin University, GPO Box U1987, Perth, WA 6845, Australia}
\author{Bahar Bidaran} 
\affiliation{ Dpto. de Física Teórica y del Cosmos, Facultad de Ciencias (Edificio Mecenas), University of Granada, E-18071, Granada, Spain}
\author{Emille E. O. Ishida} 
\affiliation{Universit\'e Clermont Auvergne, CNRS/IN2P3, LPC, F-63000 Clermont-Ferrand, France}
\author{Alberto Krone-Martins} 
\affiliation{Donald Bren School of Information and Computer Sciences, University of California, Irvine, CA 92697, USA}
\author{Aarya A. Patil}
\affiliation{Max-Planck-Institut f\"ur Astronomie, K\"onigstuhl 17, D-69117 Heidelberg, Germany}
\author{Reinaldo R. Rosa} 
\affiliation{ Lab for Computing and Applied Mathematics, COPDT-INPE-MCTI, S.J. dos Campos, SP 12245-010, Brazil}
\author{Andressa Wille}
\affiliation{Instituto de F\'isica, Universidade Federal do Rio Grande do Sul, Porto Alegre, RS 90040-060, Brazil}
\author{Richard M. Plotkin}
\affiliation{Department of Physics, University of Nevada, Reno, NV 89557, USA}
\affiliation{Nevada Center for Astrophysics, University of Nevada, Las Vegas, NV 89154, USA }
\author{Behzad Tahmasebzadeh}
\affiliation{Department of Astronomy, University of Michigan, 1085 S. University Ave., Ann Arbor, MI 48109, USA}
\affiliation{Department of Astrophysics and Planetary Science, Villanova University, 800 East Lancaster Ave., Villanova, PA 1808 }
\author{for the COIN collaboration}




\begin{abstract}
    
Tidal Disruption Events (TDEs) have long been hypothesized as valuable indicators of black holes, offering insight into their demographics and behaviour out to high redshift. TDEs have also enabled the discovery of a few Massive Black Holes (MBHs) with inferred masses of $10^4$--$10^6\,M_\odot$, often associated with the nuclei of dwarf galaxies or ultra-compact dwarf galaxies (UCDs). Here we present AT2022zod, an extreme, short-lived optical flare in an elliptical galaxy at $z=0.11$, residing within 3\,kpc from the galaxy’s centre. Its luminosity and $\sim$30-day duration make it unlikely to have originated from the host galaxy’s central supermassive black hole (SMBH), which we estimate to have a mass of $\sim 10^8\,M_\odot$. Assuming that the emission mechanism is consistent with known observed TDEs, we find that such a rapidly evolving transient could either be produced by a MBH in the intermediate-mass range or, alternatively, result from the tidal disruption of a star on a non-parabolic orbit around the central SMBH. We suggest that the most plausible origin for AT2022zod is the tidal disruption of a star by a MBH embedded in a UCD. As the Vera C.\ Rubin Observatory’s Legacy Survey of Space and Time comes online, we propose that AT2022zod serves as an important event for refining search strategies and characterization techniques for intermediate-mass black holes.

\end{abstract}


\section{Introduction}\label{sec1}

Tidal disruption events (TDEs), first proposed as a theoretical concept in the late 1970s \citep{Hills1975}, occur when a black hole shreds a passing star \citep{Frank1976,Carter1983,Rees1988, Gezari21}. The event produces optical flaring and strong broad emission lines \citep{Strubbe09}. So far, around a hundred TDEs have been observed. The majority of TDE discoveries happened in optical wavelengths, primarily by all-sky optical surveys like ASAS-SN \citep{2014ApJ...788...48S, 2017PASP..129j4502K} and the Zwicky Transient Facility \cite[ZTF,][]{ztf2019, 2019PASP..131g8001G, 2019PASP..131a8003M, 2020PASP..132c8001D}. 

Because of these features, searching for TDEs has also been suggested as a means to find intermediate mass black holes (IMBHs); whose mass is considered to lie within $10^2<M_\odot<10^6$, \citep{Strubbe09}. IMBHs are important cosmological objects as they may seed supermassive black holes \cite[SMBHs, ][]{banados18} and may also contribute to heating processes in the early Universe, in particular at epochs when X-ray ionization by compact objects plays a significant role \citep{2025ApJ...989...57N}. However, finding evidence of their existence is challenging. Besides, events that may be associated to IMBHs are difficult to classify as such. Indeed, all the current IMBH TDE candidates fall into the Massive Black Hole range (MBHs; $10^{4}-10^6 M_\odot$) and many of these candidates have alternate explanations that are equally plausible \cite[e.g.,][]{Oh25}.  

In this context, it is important to recall the signatures of previously observed IMBH TDE candidates, such as AT2020neh, which was optically selected and showed a high peak luminosity and swift rise time. Two other candidates have been identified in X-rays \citep{Lin18,Jin25}. Based on their optical counterparts, all three events were posited to be the central MBH of a dwarf galaxy, or its  stripped nucleus \citep{Lin18,Angus22,Jin25}. These three detections imply that IMBHs form and behave like scaled down SMBHs, which is very important for understanding the growth and evolution of SMBHs, particularly those found in the early Universe, and to understand the properties of their host galaxies.

A short lived optical flare, AT2024tvd \citep{2025ApJ...985L..48Y} suggests the presence of ``wandering SMBHs'' \citep{Bellovary10}, in a host galaxy with a heavier central SMBH \cite[][; see simulations by \cite{2018ApJ...857L..22T, 2021MNRAS.503.6098R,2021ApJ...916L..18R}]{2021MNRAS.505.5129B, 2021ApJ...917...17G}. This suggests that many MBHs, along with their nuclear star clusters, are acquired by galaxies early on, 
and that they will be retained at high galactocentric radii regardless of galaxy morphology or age.

 The threshed nuclear remains of larger galaxies, i.e., MBHs which retain their nuclear star clusters and are embedded in massive elliptical galaxies are called ultra-compact dwarf galaxies (UCDs; \cite{2020ApJS..250...17L}) and have been found in the Local Volume, and within the Milky Way \citep{2014Natur.513..398S, 2024Natur.631..285H, 2025ApJ...991L..24T}. Their MBHs do not appear to be interacting with stars in a manner to produce electromagnetic emission in the form of X-ray or radio, as any gas will be stripped out of the cluster within a few Myr \citep{2020MNRAS.493.1306C}. Thus, the only alternate route to observe MBHs in UCDs is through TDEs. Simulations by \cite{2025ApJ...990L..69G} suggest that between 0.1\%-10\% of observed TDEs may be due to MBHs from UCDs.

Here we present AT2022zod, a short-lived optical flare from an elliptical galaxy at $z$= 0.11, and arguments for why it appears unusual compared to the current population of identified TDEs, and show that it may be a new addition to the category of UCD MBHs. 
 We further provide suggestions for how to improve science outcomes for fast rising TDEs that will be newly discovered in the era of the Vera C. Rubin Observatory \citep{2019ApJ...873..111I}.

\begin{figure}
    \centering
    \includegraphics[width=\linewidth]{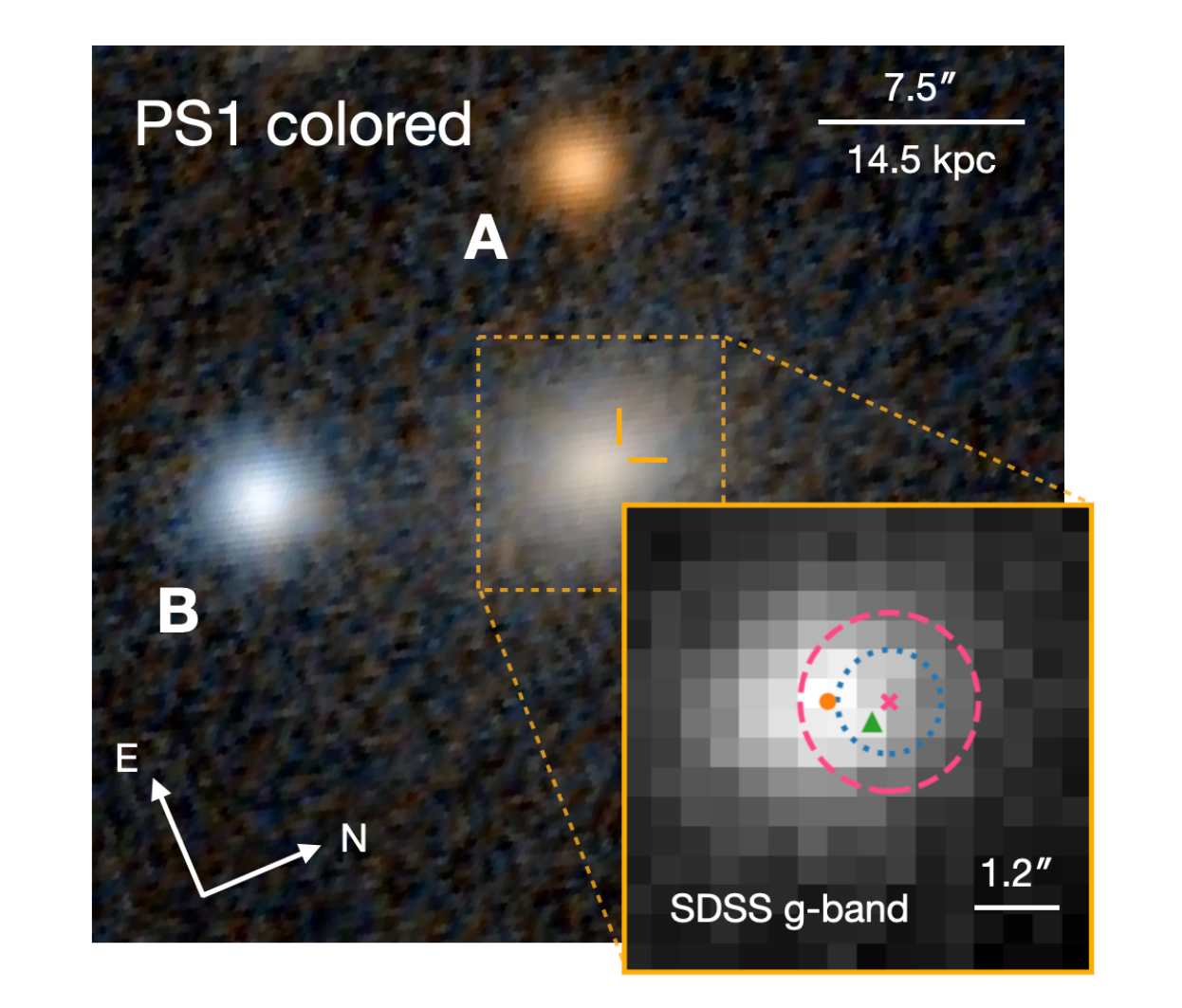}
    \caption{Position of the transient in its host galaxy. The main panel shows a PS1 colored image of the galaxy, while the inset shows the SDSS image in $g-$band. The most nearby objects labeled with A and B are foreground stars Gaia DR3 850874871499596288 and Gaia DR3 850874772715782912 (with measured parallaxes of 3.81 mas and 0.69 mas respectively). The pink cross marks the position of the transient, with the blue dotted circle showing the 0.7" positional uncertainty reported, plus a pink dashed circle showing a more conservative positional uncertainty at 1.2". The green triangle  marks the SDSS coordinates of the galaxy and the orange circle marks the centroid position of the galaxy, estimated with a 2D Gaussian fit. At z=0.11, 1" is equivalent to 2 kpc. This suggests that the transient may be coincident with the galaxy center, or up to a few kpc outside of it.  }
    \label{fig:placeholder}
\end{figure}

\section{Methods}\label{sec2}

We performed a systematic search for flaring events occurring in elliptical galaxies by cross-matching alerts generated by  the ZTF, between November 2019 and August 2025, with elliptical galaxies identified by Galaxy Zoo Data Release 1 \citep{2011MNRAS.410..166L}. This was performed  using  the online cross-matching tool\footnote{\url{https://ztf.fink-portal.org/xmatch}} developed by the Fink broker \citep{moller2021}. Once interesting candidates were identified, full ZTF Data Release 23 (DR23) light curves were acquired using the SNAD API \citep{malanchev2023} and the Infra-Red Service Archive\footnote{\url{https://irsa.ipac.caltech.edu/docs/program_interface/ztf_lightcurve_api.html}} (IRSA). 

A paper summarizing this search and estimating rate limits is currently in preparation. Here, we report the best candidate found in this search. AT2022zod is an unclassified event originally discovered by ZTF at $g = 19.2$ mag on 2022-10-18 and reported as ZTF18aarlhfw \citep{2022TNSTR3230....1F}. We emphasize that AT2022zod was neither classified as a TDE, nor as a SNe. 

\begin{figure}
    \centering
    \includegraphics[width=0.99\linewidth]{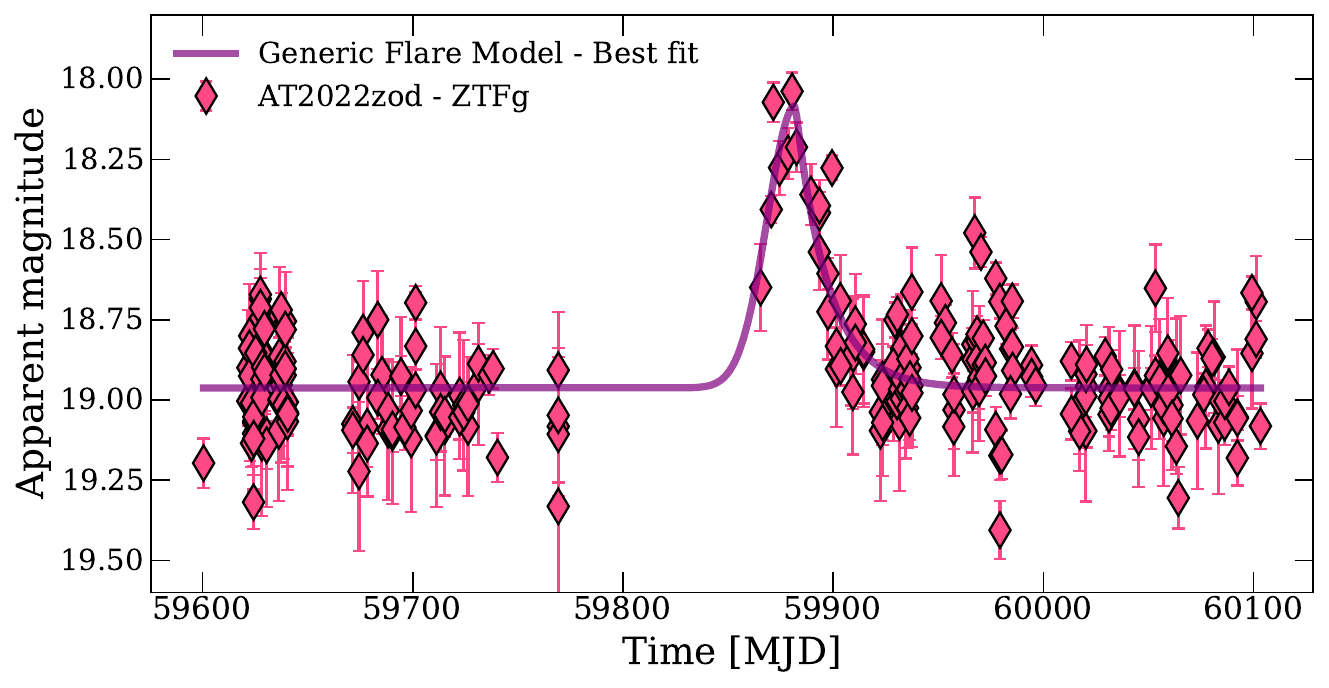}
    \caption{Light-curve data and best-fit flare model for AT2022zod in the ZTF \textit{g} band. The points show the observed apparent magnitudes with their uncertainties, and the solid line shows the Gaussian-rise, exponential-decay model parameterized by $r_{0}$, $A$, $t_{0}$, $t_{g}$, and $t_{e}$ (best-fit parameter values are shown in Table \ref{tab:ztf_fit_params}). Time is given in Modified Julian Date (MJD).}
    \label{fig:zod_LC}
\end{figure}

We analyse ZTF lightcurves of the optical flare AT2022zod, which lasted roughly 30 days between Oct 13, 2022 and Nov 18, 2022 (Fig. \ref{fig:zod_LC}). It is hosted by elliptical galaxy SDSS J105602.80+561214.7, at z=0.11 (Fig. \ref{fig:placeholder}).  We explore a large range of possible scenarios for the origin of this flare.

\subsection{The Host SDSS J105602.80+561214.7}

\begin{figure}
  \centering
  \includegraphics[width=\linewidth]{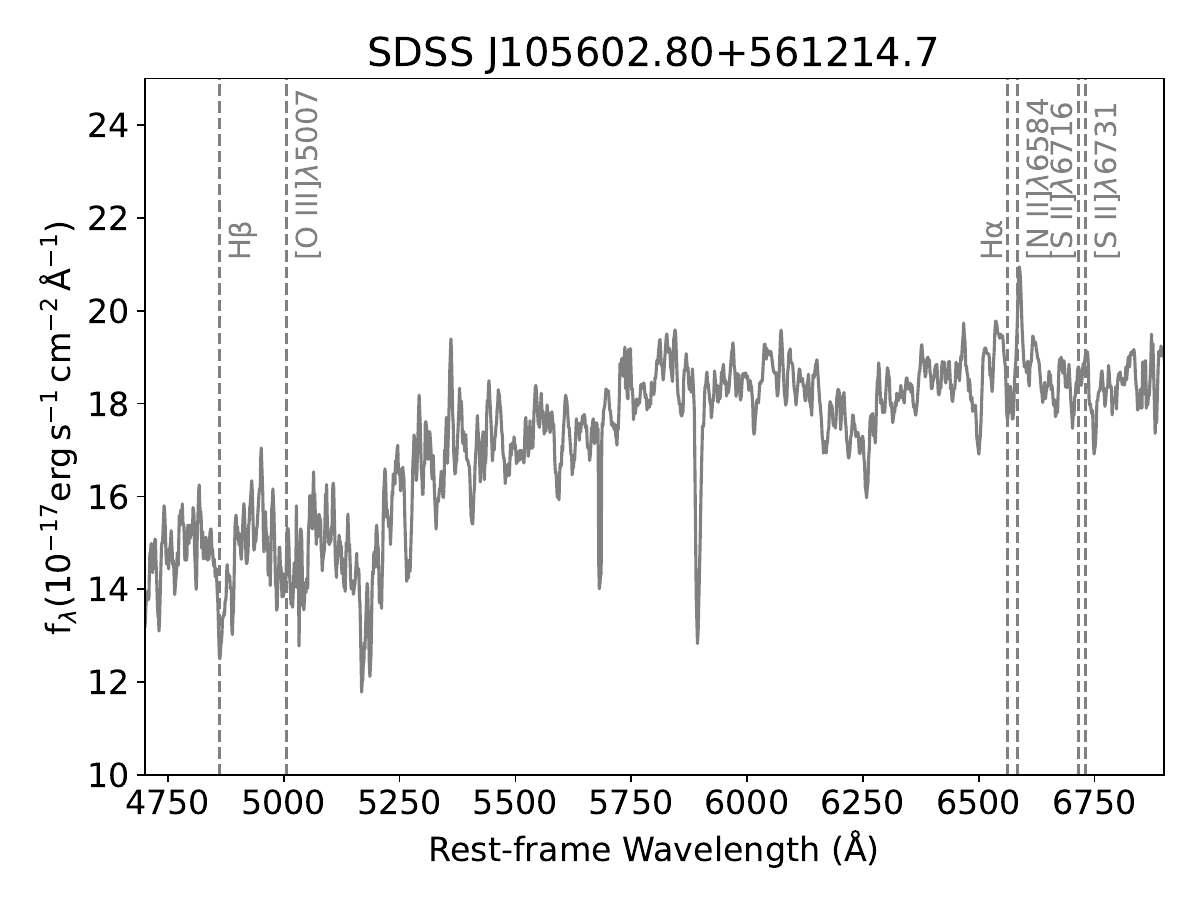}
\caption{Spectrum of SDSS J105602.80+561214.7. Dashed lines show expected strong emission lines at the redshift of the galaxy in case of AGN activity.}
  \label{fig:hfw_host_spectrum}
\end{figure}

\begin{figure}
    \centering
    \includegraphics[width=\linewidth]{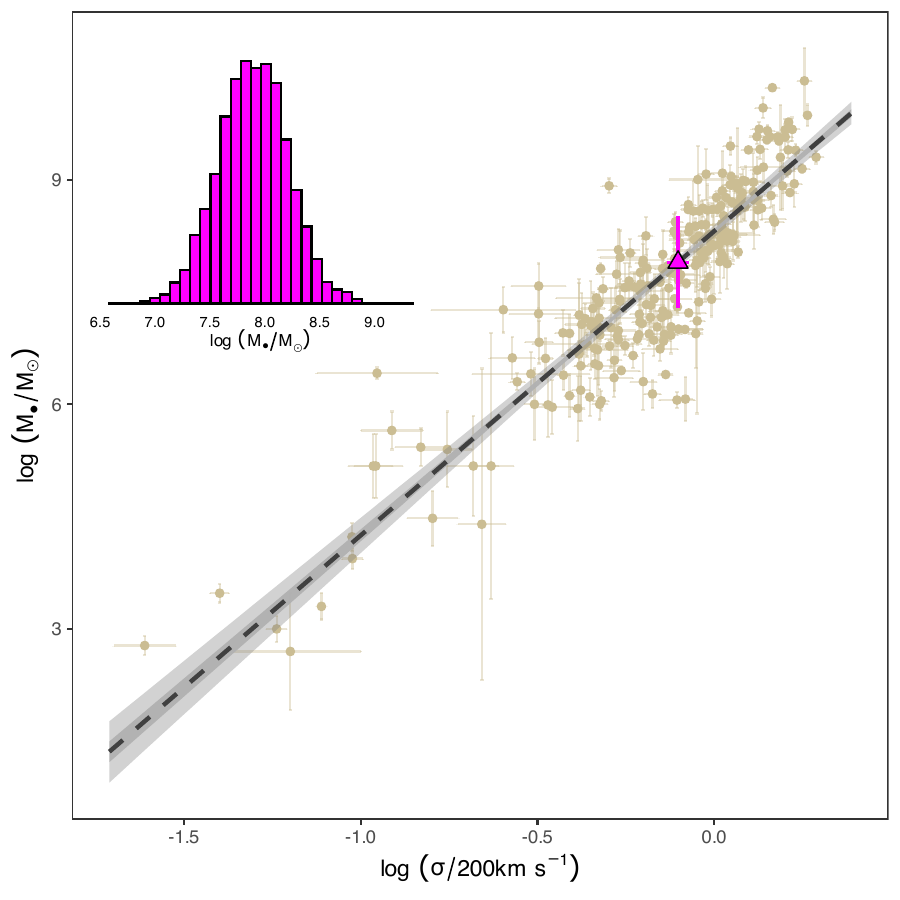}
    \caption{M--$\sigma$ relation. The data points are from the compilation of \cite{vdb2016}.
  The central black hole of 105602.80+561214.7 is closer to the high mass end of the M-sigma relation, than to where IMBHs are expected to be.}
    \label{fig:msigma}
\end{figure}

The elliptical galaxy was observed with the SDSS 2.5-m telescope on 9 April 2002 \citep{2000AJ....120.1579Y}. Its spectrum (see Fig.~\ref{fig:hfw_host_spectrum}) lacks prominent emission lines; Balmer lines (H$\alpha$, H$\beta$) and key forbidden lines ([O III] $\lambda$5007, [N II] $\lambda$6584, [S II] $\lambda$$\lambda$6717, 6731) are at best detected at very low signal-to-noise. Based on the measured stellar velocity dispersion of 
$158.08 \pm 11.73\,\text{km\,s}^{-1}$, 
we infer, through a Bayesian hierarchical model \citep{hilbe2017}, 
the presence of a SMBH with a mass of 
$M_\bullet \simeq 1.0 \times 10^{8}\,M_\odot$, 
and a 95\% credible interval spanning 
$9.7 \times 10^{6}$ to $2.6 \times 10^{8}\,M_\odot$ (Fig.~\ref{fig:msigma}). 
This estimate is obtained using the standard 
$M_\bullet$--$\sigma$ relation 
\citep{2002ApJ...574..740T, 2013ApJ...764..184M, 2013ARA&A..51..511K}, 
and is consistent with the range expected from these empirical calibrations.

Assuming the \cite{taylor11} empirical relation for $z < 0.65$

\begin{equation}
\log_{10}\!\left(\frac{M_\star}{L_i}\right)
 = -0.68 + 0.70\,\bigl[(g - i)\bigr],
\end{equation}

\noindent and the SDSS magnitudes g= 17.90 and i=16.364 we estimate a stellar mass of

\begin{equation}
M_\star \approx 8.5 \times 10^{10}\, M_\odot.
\end{equation}

\noindent Adopting the Planck 2018 cosmology \citep{2020A&A...641A...6P}, we obtain a luminosity distance of 526.86 Mpc for the source.

Galaxies of this class will contain, nuclear star clusters (NSCs), the densest stellar systems in the Universe \citep{Neumayer20}. Two channels can explain their formation: i) ~in situ formation scenario, where the NSC forms at the galaxy centre from infalling gas and local star formation; ii)~
gas-free accretion of globular clusters (GCs) that spiral inwards due to dynamical friction.
The first channel dominates at high host galaxies stellar masses \cite[][ $M_{\mathrm{gal}} \geq 10^{9}\, M_{\odot}$]{fahrion21, fahrion22}, including the host galaxy of AT2022zod.

\begin{figure*}

\includegraphics[scale=0.60]{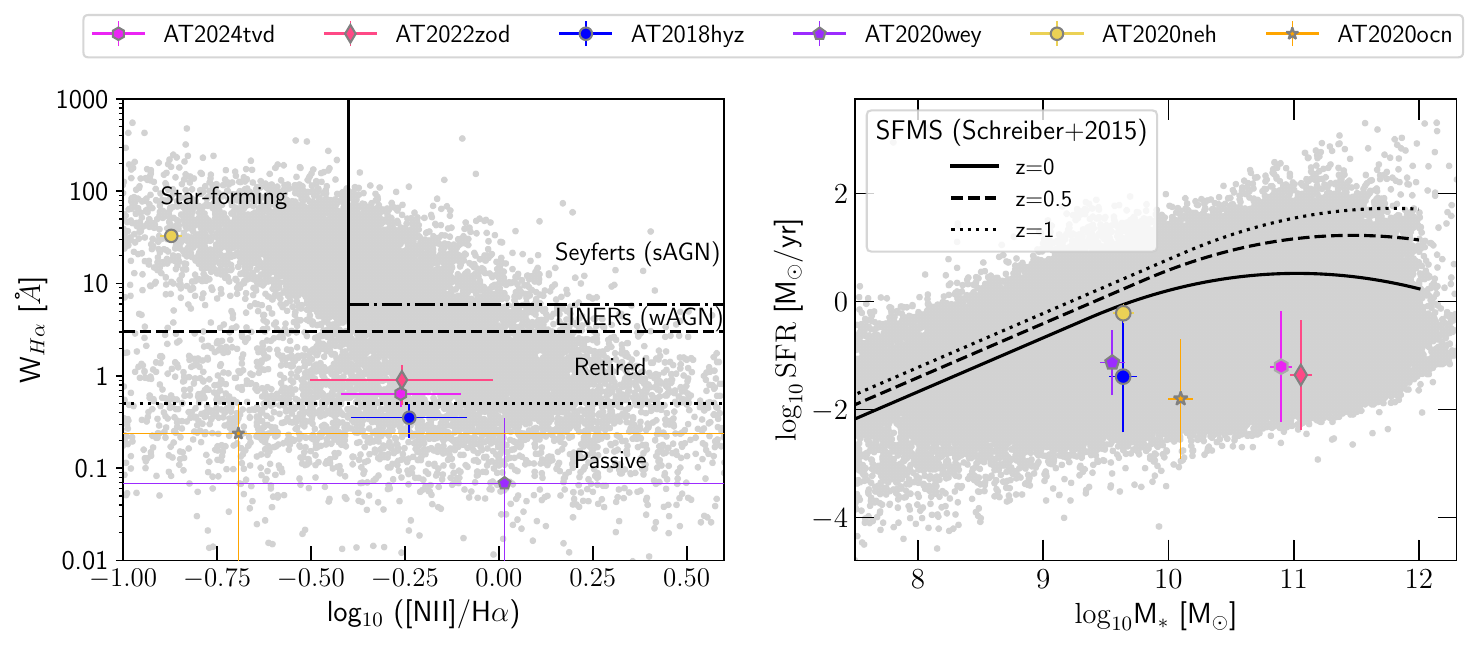}

\caption{\textbf{[Left]}: AGN classification diagnostic (WHAN; \cite{CidFernandes2011}), showing where the \cite{2023ApJ...942....9H} sample of TDEs with available SDSS spectroscopy fall. AT2022zod and AT2024tvd's host galaxies are classified as retired galaxies. \textbf{[Right]}: Star formation of hosts for AT2022zod and other known TDE host galaxies as indicated in the legend. AT2022zod and AT2024tvd have very low specific star formation rates, unlike the hosts of many observed TDEs, found in post-starburst galaxies \citep{2024ApJ...960...69W}. A random subsample of 50,000 galaxies in the SDSS value-added catalogue MPAJHU-DR8 are shown by the grey points in both panels to avoid overcrowding.}
\label{Fig:sfms}
\end{figure*}

Based on its position in the WH$\alpha$ versus [NII]/H$\alpha$ (WHAN; \cite{CidFernandes2011}) emission-line diagnostic diagram \cite[see Fig.\ref{Fig:sfms},][]{CidFernandes2011}, we conclude that the host galaxy of AT2022zod is a retired galaxy.

\begin{figure}
    \centering
    \includegraphics[width=0.9\linewidth]{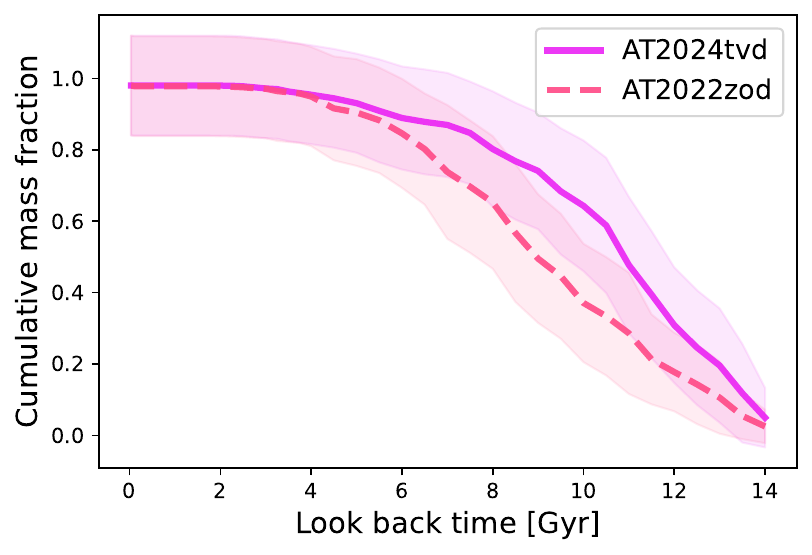}
    \caption{Host galaxy star formation histories, based on SDSS spectra, of  AT2022zod (dashed line) and AT2024tvd (solid line), showing that the latter host appears to have ceased star formation a few Gyr earlier than AT2022zod.  }
    \label{fig:bahar}
\end{figure}

We derived the star formation histories (SFHs) of AT2024tvd and
AT2022zod using optical spectra from the 7th data release of the Sloan
Digital Sky Survey (SDSS-DR7), obtained with the 2.5 m telescope at the
Apache Point Observatory. The spectra have a resolving power of R $\approx$ 1500
at $\lambda$ = 3800 ${\AA}$ and were collected using a 3-arcsecond fiber, providing
integrated light within this aperture. After correcting for Galactic
foreground extinction, assuming R$_{(v)}$ = 3.1, the \cite{1989Cardelli}
extinction law, and E(B-V) values corresponding to each galaxy’s
coordinates, we fit the spectra using the pyPipe3D full spectral-fitting
pipeline \citep{2016Sanchez, 2022Lacerda}. For the fits, we adopted 1271
single stellar population (SSP) models based on the MILES stellar
library \citep{Sanchez-Blazquez2006, Cenaro2007, Falcon-Barroso2011}.
These SSPs are built using the BaSTI isochrones \citep{Pietrinferni2009}
and a bimodal initial mass function (IMF) with a slope of 1.3
\citep{Vazdekis1996}. The models span ages from 0.03 to 14 Gyr and
metallicities from $-$2.27 $<$ [M/H] $<$ +0.40. The fits were performed over
the wavelength range 400–690 nm, which includes the principal stellar
absorption features, such as H$\beta$, that constrain stellar age and
metallicity.
Following the approach of \cite{2025ABidaran}, we carried out 50
independent Monte Carlo (MC) realizations for each galaxy. For each
realization, we constructed the cumulative mass fraction (CMF), derived
from the stellar mass formed as a function of look-back time based on
the full spectral-fitting outputs. In Fig.~\ref{fig:bahar}, we present the mean
CMF across the 50 realizations as a solid line, with the associated
uncertainty, quantified as the standard deviation among the
realizations, shown as a shaded region.
\\
\\
Given that AT2022zod and AT2024tvd, along with several other well-studied TDEs, lie in host galaxies with available SDSS spectra, we can contruct the WHAN diagram, an emission-line diagnostic for distinguishing the dominant ionizing source in galaxies \citep{CidFernandes2011}. To construct the diagram, as shown in the left panel of Fig. \ref{Fig:sfms}, we use emission line fluxes from the value-added catalogue of the eighth data release of (SDSS) DR8 \citep{2011SDSSDR8} compiled by the group at Max Planck Institute for Astrophysics and the John Hopkins University (MPA--JHU; \cite{Brinchman2004}) and the Portsmouth Group \citep{2013Thomas_sdss}. W(H$\alpha$) serves as a proxy for the strength of the ionizing radiation field, while [NII]/H$\alpha$ traces the hardness of the ionizing spectrum and gas-phase metallicity. Compared to traditional BPT diagnostics \citep{BPT1981}, the WHAN diagram is applicable to a broader range of spectra, including systems with weak emission lines, making it particularly useful for low-luminosity galaxy samples. The WHAN diagram with the host galaxies of AT2022zod, AT2024tvd and a few other known TDEs are shown in Fig.
\ref{Fig:sfms}. MPA--JHU DR8 also contains estimates of the galaxy star formation rates (SFRs) calculated using H$\alpha$ luminosity \citep{Brinchman2004} and stellar mass values estimated as described by \cite{2003Kauffman}, both calculated assuming a \cite{Kroupa2001} initial mass function (IMF). In the right panel of Fig. \ref{Fig:sfms}, we place the host galaxy of AT2022zod along with some other TDE hosts on the star-forming main sequence (SFMS), using the relation from \cite{Schreiber2015}, calculated at redshift values of 0, 0.5 and 1 as a reference. The host of AT2022zod lies well below the SFMS, confirming its quiescent nature and strongly sub-main-sequence specific star-formation rate.


The host galaxy of AT2022zod lies close to the boundary of being a passive ``line-less" galaxy, similar to the position of the AT2024tvd host. Points below the dotted line, such as host galaxies of AT2020wey and AT2020ocn, exhibit highly uncertain emission-line ratios due to the weak signal-to-noise of the emission lines. When placed on the star-forming main sequence (see supplemental material and the right panel of Fig. \ref{Fig:sfms}), the system is clearly consistent with a quiescent, non–star-forming galaxy, with no evidence for ongoing star formation that could mimic weak AGN-like line ratios. The combination of its low H$\alpha$ equivalent width and its strongly sub-main-sequence specific star-formation rate (sSFR = SFR/M$_*$) favors an interpretation in which the nebular emission is dominated not by an active nucleus but by ionization from evolved stellar populations. In turn, this also allows us to rule out the presence of an active galactic nucleus (AGN) more robustly (see sections \ref{ruling_out_nontde} and \ref{agn_flaring}) and strengthens the case that AT2022zod is not associated with persistent AGN-like nuclear activity. The similarity in position on the WHAN diagram, low sSFR, and other host galaxy properties between AT2022zod and the well-studied AT2024tvd host supports the notion that both transients originated from BHs in low-sSFR, quiescent galaxies. This reinforces the emerging pattern that TDEs in such systems could potentially provide a unique window into the MBH population.

Although the host galaxies of AT2022zod and AT2024tvd have comparable masses, the host of AT2024tvd quenched its star formation earlier than that of AT2022zod (Fig. \ref{fig:bahar}), implying a distinct evolutionary history. Even so, both galaxies are unusual environments for such short-lived TDEs.

\subsection{Multiwavelength Observations}

SDSS J105602.80+561214.7 has not been observed in X-ray by \textit{Swift, XMM-Newton} or \textit{Chandra}. Its position is not in the eROSITA-DE sky region, and no Fermi 4FGL sources were found at the position of the galaxy. The event did not occur during any LIGO-Virgo-Kagra runs \citep{LVK2020CQGra..37e5002A}. 

However, the galaxy was by chance observed by the Very Large Array Sky Survey \citep{2020PASP..132c5001L} just 5 months after the event. No radio counterparts were observed above a non-detection threshold of $3\times 10^{38}$ erg/s. This upper limit suggests no high levels of current star formation or major activity of the central SMBH.

\subsection{AT2022zod Observational Properties}

To assess whether AT2022zod is statistically consistent with a transient-like event rather than baseline stochastic variability, we modeled its light curve using a generic flare profile consisting of a Gaussian rise followed by an exponential decay. The observational properties (Table~\ref{tab:ztf_fit_params}) were estimated using the ZTF $g$-band light curve (Fig. \ref{fig:zod_LC}) over a 400-day interval centered on the epoch of maximum brightness (MJD $\simeq$ 59880; 2022 October 28). The event exhibits a total duration of $t_{\mathrm{dur}} \simeq 28.7$~days, with a Gaussian rise timescale of $t_g = 13.24 \pm 3.68$~days and an exponential decay timescale of $t_e = 15.44 \pm 2.38$~days. The flare reached its maximum flux at $t_0 = 59881.996 \pm 2.124$~MJD (2022~October~29). The optical brightness, in the g-band, increased at an average rate of $\sim 0.66~\mathrm{mag~day^{-1}}$ during the rise and declined at a rate of $\sim 0.057~\mathrm{mag~day^{-1}}$ thereafter.

\begin{table*}
\centering
\caption{Best–fit parameters for AT2022zod using the Gaussian–rise, exponential–decay model in the $g$ and $r$ bands. 
Parameters are: baseline magnitude ($r_0$), flare amplitude ($A$), peak epoch ($t_0$), Gaussian rise timescale ($t_g$), and exponential decay timescale ($t_e$). Uncertainties are $1\sigma$.}
\label{tab:ztf_fit_params}
\begin{tabular}{lcc}
\hline\hline
Parameter & ZTF-$g$ & ZTF-$r$ \\
\hline
$r_0$ & $18.962 \pm 0.010$ & $17.865 \pm 0.008$ \\
$A$ & $0.880 \pm 0.079$ & $0.329 \pm 0.046$ \\
$t_0$ (MJD) & $59881.996 \pm 2.124$ & $59871.344 \pm 5.113$ \\
$t_g$ (days) & $13.236 \pm 3.681$ & $3.927 \pm 3.866$ \\
$t_e$ (days) & $15.442 \pm 2.384$ & $57.056 \pm 11.066$ \\
Total Duration ($t_g$+$t_e$) (days) & $28.678 \pm 6.065$ & $60.983 \pm 18.859$ \\
\hline
\end{tabular}
\end{table*}

\subsection{Comparison with Observed TDEs}

\begin{figure*}
    \centering
    \includegraphics[width=6in]{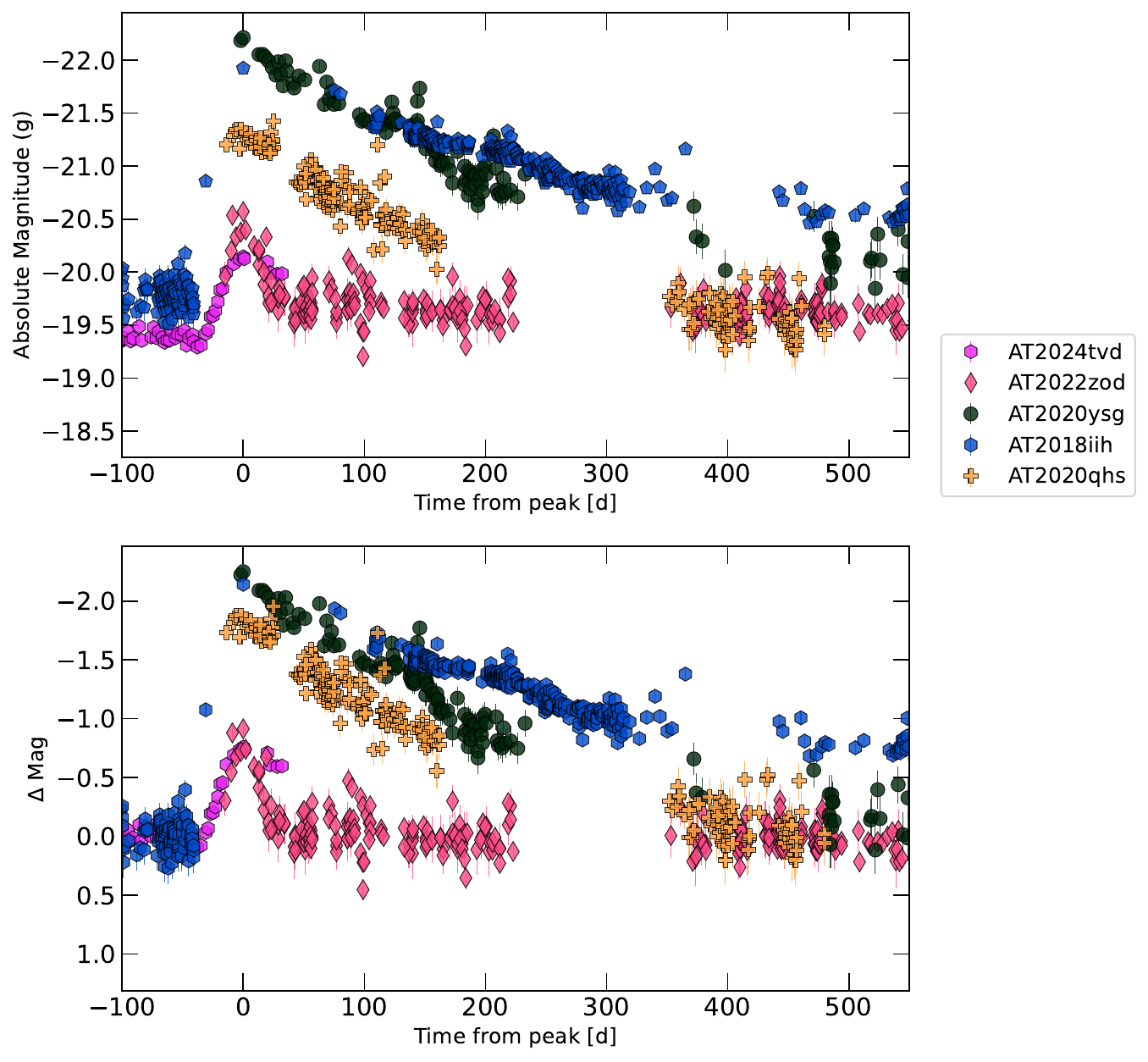}
    \caption{Comparison of TDEs hosted by the 5 most massive galaxies. AT2022zod and AT2024tvd are much shorter lived and less luminous than other events, suggesting that neither of these events are produced by the central SMBH of their host galaxies, which are both around $10^8 M_\odot$.}
    \label{fig:bigbois}
\end{figure*}

\begin{figure}
    \centering
    \includegraphics[width=3.65in]{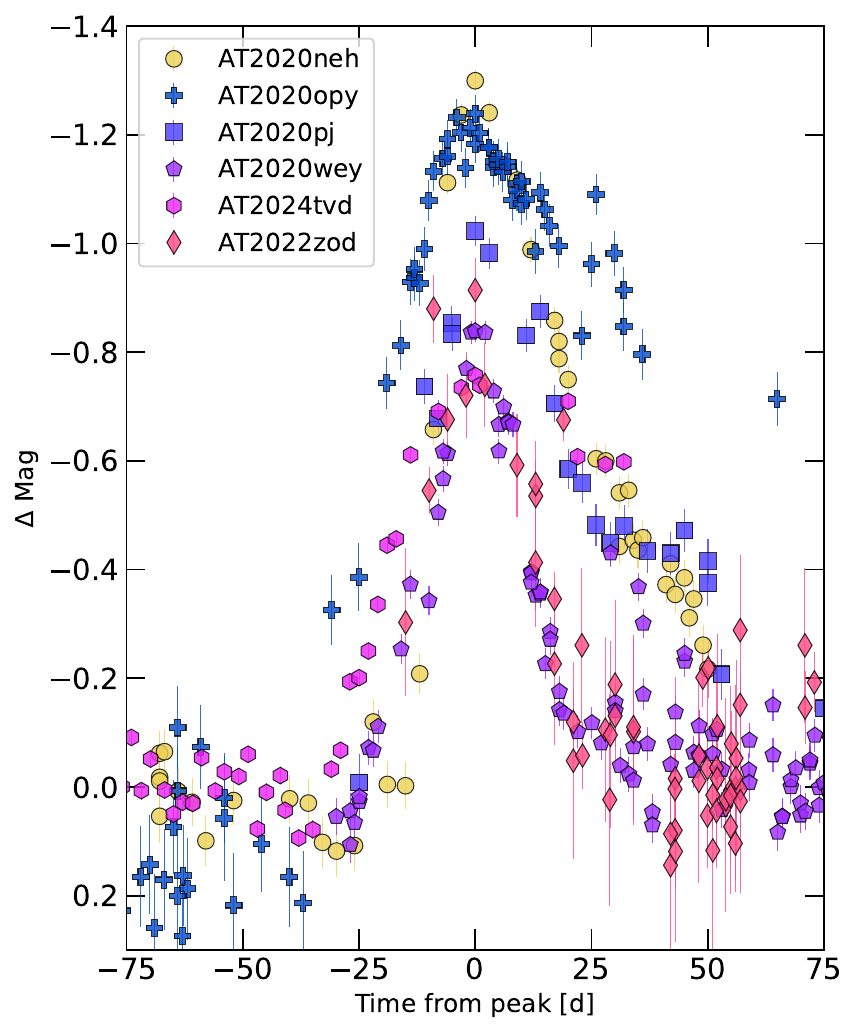}
    \caption{Comparison of observed lightcurve scaled to host galaxy mass for TDEs which achieve similar absolute magnitudes to AT2022zod. AT2020wey, AT2022zod appear to have the most similar shape, although AT2022zod's peak is 2.5 orders of magnitude higher.  }
    \label{fig:deltamag}
\end{figure}

We find that AT2022zod is at a reasonably high redshift, and is hosted by an extremely massive galaxy, which obscures more of the lightcurve, compared to other TDEs hosted by lower mass or more nearby galaxies. The available data therefore likely capture only the upper portion of the event’s light curve, due to being masked by the host galaxy, resulting in less complete temporal coverage than is typical for similar transients. However, when comparing with any known TDE with a host galaxy mass within 0.5 dex within AT2022zod, we see that both AT2024tvd and AT2022zod are shorter lived and less luminous, by a factor of 10 (Fig. \ref{fig:bigbois}). However, AT2022zod reaches a higher peak luminosity within a shorter timescale than AT2024tvd, implying that AT2022zod may be powered by a less massive black hole than AT2024tvd. If we compare AT2022zod and AT2024tvd to the sample of TDEs hosted by lower mass galaxies, which reach similar luminosities above the host galaxy background,  we find that the observed timescales of AT2022zod are most consistent with AT2020wey, although AT2020wey's peak luminosity is roughly 2.5 orders of magnitude lower than that of AT2022zod (Fig. \ref{fig:deltamag}).

\begin{figure*}
    \centering
    \includegraphics[width=1\linewidth]{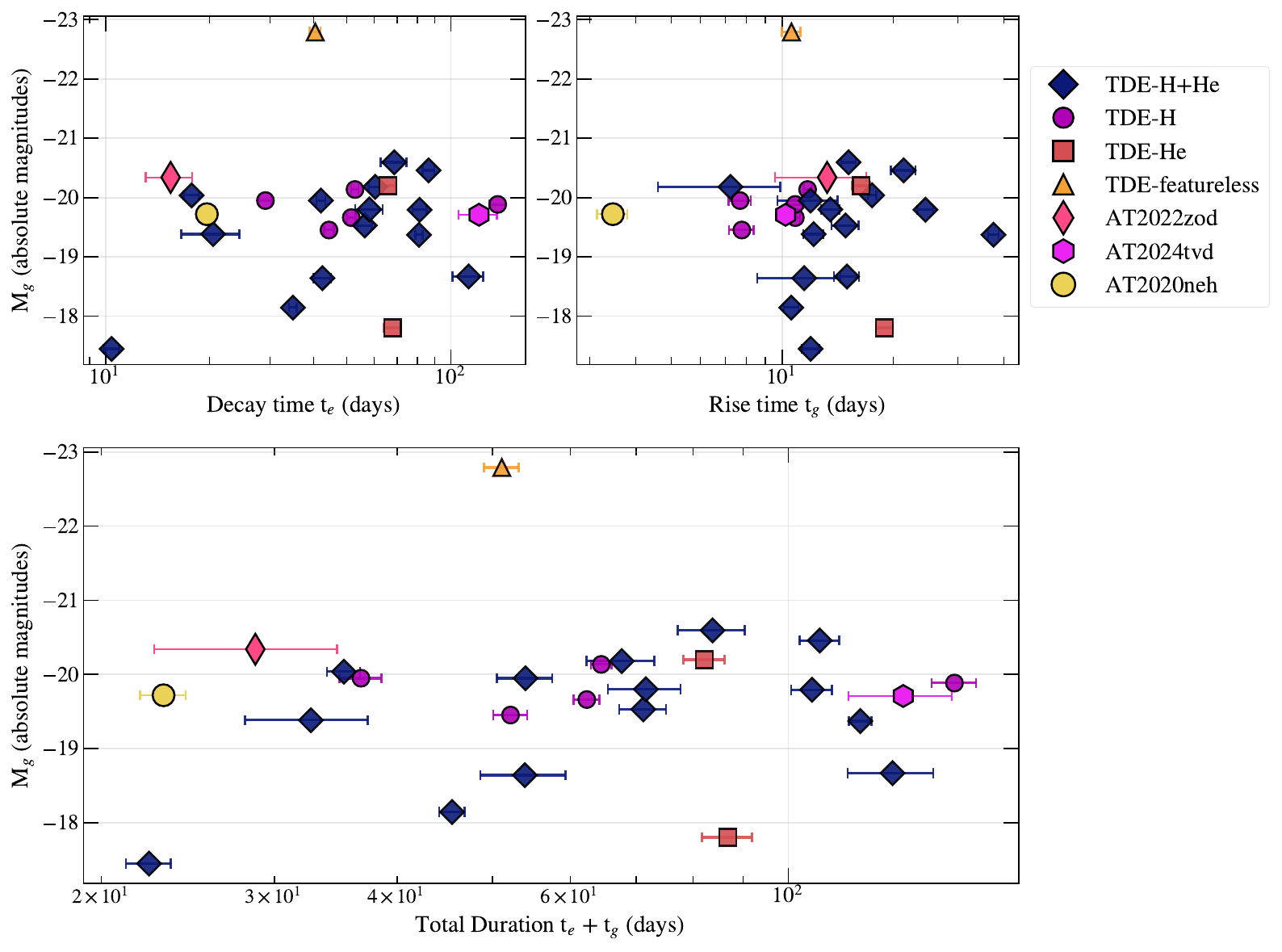}
 \caption{Observed TDE length versus peak absolute magnitude. We see that AT2022zod has a much shorter observed duration for a high peak magnitude, compared to the other TDEs with similar peak magnitudes. However, we note that this may be a combination of an intrinsic difference in the TDE, or because a combination of a relatively fainter event occuring at higher redshift in a higher mass galaxy are obscuring more of the lightcurve than in the comparison systems.}
    \label{Fig:rise_decay_mag}
\end{figure*}

We examine the observed timescales of AT2022zod  against the observed timescales of the TDE sample from \cite{2023ApJ...942....9H}. These events are color-coded by spectroscopic subtype (H, He, H+He, and featureless), following the classifications in \cite{2023ApJ...942....9H}. For context, we also include the candidate MBH TDEs AT2024tvd \citep{2025ApJ...985L..48Y} and AT2020neh \citep{Angus22}. All parameters shown in Fig. \ref{Fig:rise_decay_mag} were obtained by fitting a Gaussian flare model to the observed light curves.

\begin{figure*}
    \centering
    \includegraphics[width=1\linewidth]{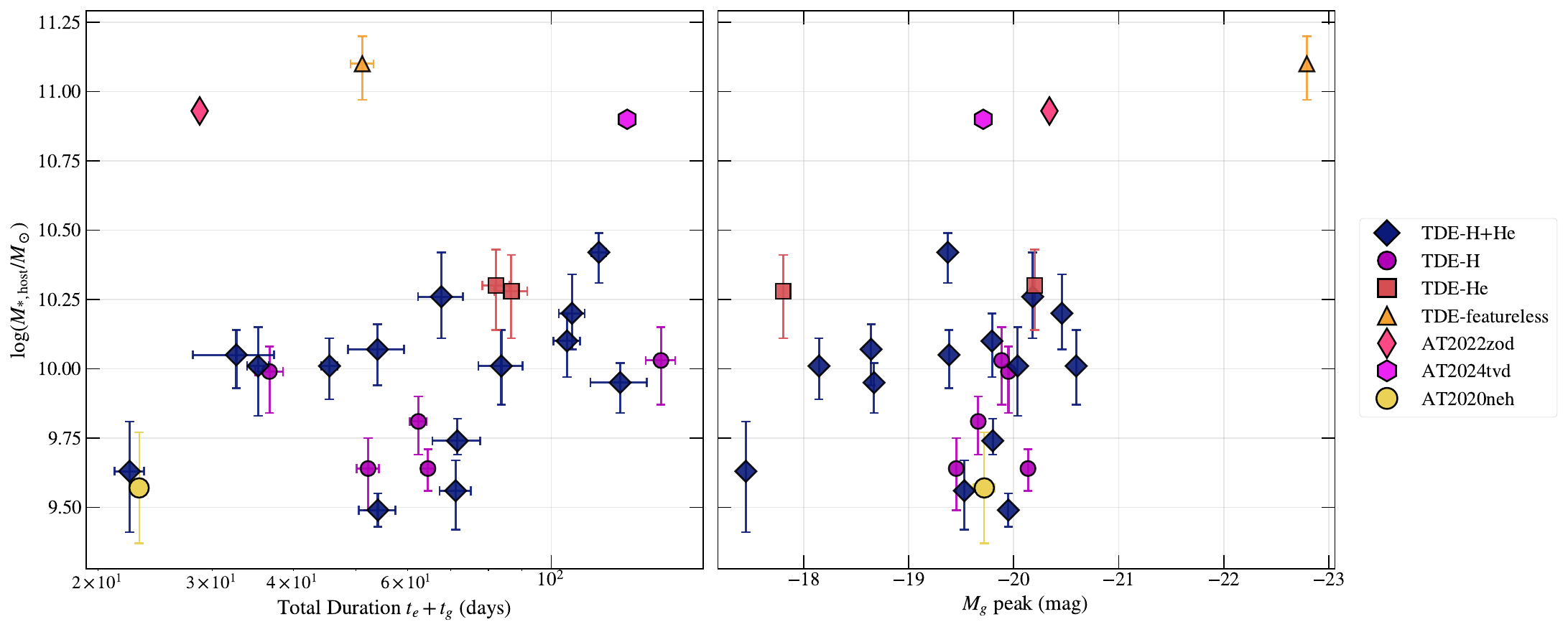}
    \caption{Properties of TDEs (total observed duration, peak magnitude) versus host galaxy mass. AT2022zod, AT2024tvd and AT2020riz are clear outliers in this trend, however, AT2020riz's lightcurve is best described by a significantly larger BH mass than AT2022zod and AT2024tvd \citep{2023ApJ...942....9H}.   }
    \label{Fig:exceptional}
\end{figure*}

AT2022zod exhibits one of the fastest exponential decay timescales compared to other TDEs with a similar peak magnitude, surpassed only by AT2020wey, which has $t_e = 10.4 \pm 0.5$ days but a significantly lower peak luminosity (2.5 orders of magnitude fainter than AT2022zod). We also see that AT2022zod has the shortest duration event for its host-galaxy mass (Fig. \ref{Fig:exceptional}). 
Building on these comparisons, the evidence points toward AT2022zod not being powered by the central SMBH, but instead originating from a lower-mass black hole within the system. A MBH, either residing in a stripped nuclear star cluster or in an off-nuclear compact stellar system, would naturally produce shorter characteristic timescales while still being embedded in a comparatively massive galaxy. Although further diagnostics are needed to firmly establish this scenario, the photometric evolution of AT2022zod is broadly consistent with expectations for a TDE generated by a MBH.

\subsection{Lightcurve Modeling}

We examined and systematically constrained several possible origins for the flaring event AT2022zod, including AGN variability, a supernova explosion, a compact-object merger, and the tidal disruption of a star by a SMBH. We first characterized the photometric and temporal properties of AT2022zod.

We obtained the light curve of the candidate from the ZTF DR23. Each ZTF light curve contains the Modified Julian Date (MJD), point-spread function (PSF) apparent magnitude, $1\sigma$ photometric uncertainty, and corresponding image-quality flags (\texttt{catflags}) for each observation. All observations with nonzero \texttt{catflags} (indicating poor-quality or contaminated measurements) were excluded from the analysis. We also discarded the i-band data owing to sparse temporal sampling.

The optical lightcurve produced by a TDE has different model approximations based on different physical backgrounds \cite[e.g][]{2019ApJ...872..151M, 2020ApJ...904...73R, 2023PASP..135c4102K}. For example, {\tt MOSFiT} \citep{2019ApJ...872..151M} is one of the most popular code for fitting the lightcurve of a TDE candidate assuming a reprocessing layer as the source of the optical peak. Another possibility is the {\tt TDEmass} \citep{2020ApJ...904...73R}, which assumes slow circularization where the source of the optical peak is the shock in the intersecting/colliding debris. Here, we mainly focus on {\tt TiDE} \citep{2023PASP..135c4102K} code, although we performed the fitting procedure using {\tt MOSFiT} and a simplified analytical prescription implemented in \texttt{Redback}, which assumes a constant bolometric luminosity prior to the peak followed by the canonical fallback decay $L \propto t^{-5/3}$.

{\tt TiDE} is a modular, open-source code for modeling monochromatic TDE lightcurves. It was originally written in C++, but a Python wrapper is now also available as {\tt TiDEpy}. It assumes a super-Eddington wind component as the source of the optical peak, which is a major difference compared to {\tt MOSFiT}, where the fallback accretion rate cannot exceed the Eddington limit. In {\tt TiDE}, the fallback accretion rate can be calculated using semi-analytic prescription of \cite{2009MNRAS.392..332L}, whereas {\tt MOSFiT} relies on scaled hydrodynamical simulations. These differences suggest that the best-fit parameters will be extremely model-dependent, as demonstrated by \cite{KScomp}.

\begin{figure}
    \centering
    \includegraphics[width=3.9in]{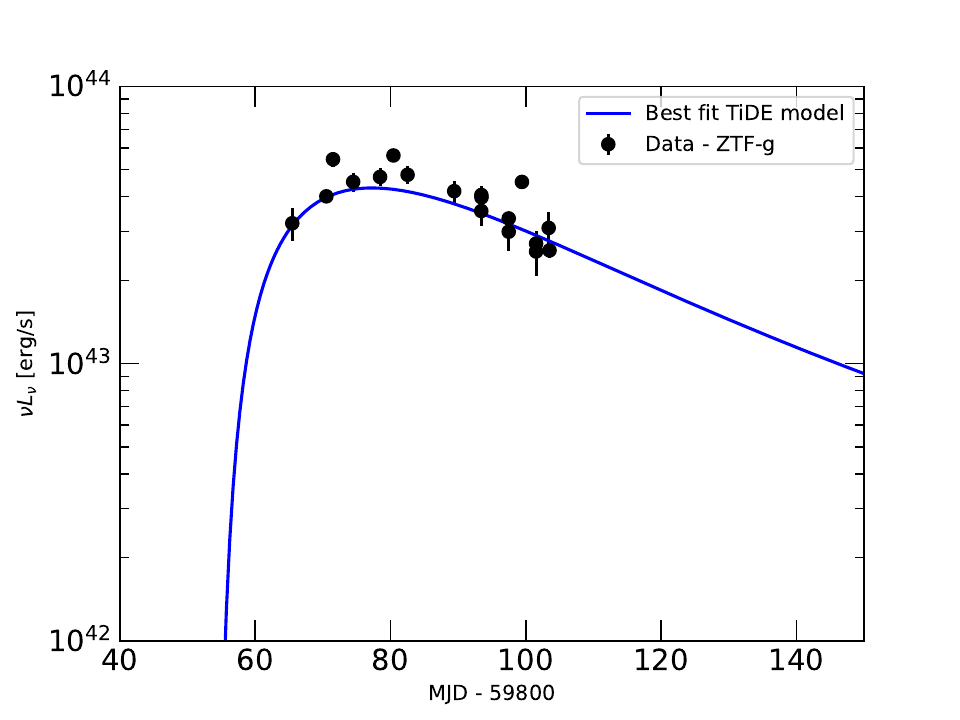}
    \caption{The optical light curve of AT2022zod (black points) compared with the best-fitting \texttt{TiDEpy} model (solid blue line). The vertical axis shows the luminosity in erg\,s$^{-1}$, and the horizontal axis shows the time in days relative to MJD\,59800.}
    \label{fig:tidepy_bestfit}
\end{figure}

The package \texttt{TiDE} \citep{2023PASP..135c4102K}\footnote{\url{https://github.com/stermzsofi/TiDE}} provides us a flexible semi-analytic TDE model to explore lightcurves produced by less massive BHs. We modeled the ZTF g-band light curve of AT2022zod using \texttt{TiDEpy}. Prior to fitting, the photometry was corrected for Milky Way extinction and converted to luminosity using the source’s luminosity distance. The fitting was performed with the \texttt{scipy} \citep{2020SciPy-NMeth} \texttt{curve\_fit} method. We tested several \texttt{TiDE} models and found that a main-sequence star with a polytropic index of $\gamma = 4/3$ provides the best agreement with the observed light curve. The corresponding parameters are listed in Table \ref{tab:tide_bestfit}, and the best-fit curve is shown in Fig. \ref{fig:tidepy_bestfit}. The fitted parameters are the same as in \cite{KScomp}, except for the $\epsilon_{\rm rep}$ parameter, which was fixed at $\epsilon_{\rm rep} = 0$ in order to achieve the best p-value. Based on this fit, the inferred black hole mass is $5.4 \times 10^5\,M_\odot$, placing it near the boundary between the IMBH and SMBH regimes.

\begin{table*}

\caption{The initial parameters for the fit, the allowed fitting ranges, the best-fit parameters obtained with {\tt TiDEpy} and its reduced $\chi^2$ and the p-value of the Kolmogorov–Smirnov test. $^\star$ Although the parameter $f_{\rm v}$ was found to be 10 during the fitting procedure, we note that its true value for the best-fit model is $5.36$, which is a consequence of the limitation that the velocity of the wind cannot exceed the speed of light.}
\begin{tabular}{l|cccc}
\hline
\hline
Parameter & Initial fitting value & Fitting range & Best-fit value\\
\hline
$t_{\rm ini}$ & $55$ & $[40.0;70.0]$ & $51.02 \pm 1.43$\\
$M_6$ & $0.1$ & $[0.001;40.0]$ & $0.54 \pm 0.05$\\
$m_*$ & $10.0$ & $[1.0;40.0]$ & $22.30 \pm 1.59$\\
$\eta$ & $0.008$ & $[0.0001;0.4]$ & $ 0.02 \pm 0.0007$\\
$f_{\rm v}$ & $5.0$ & $[1;10]$ & $5.36^{\star}$ \\
$t_{\rm diff}$ & $10.0$ & $[0.0;60.0]$ & $9.99 \pm 1.55$\\
Reduced $\chi^2$ & -- & -- & $9.43$\\
KS-test $p$ value & -- & -- & $0.29$\\
\hline
\end{tabular}

\label{tab:tide_bestfit}
\end{table*}





\begin{table}
\caption{Best-fit MOSFiT parameters for AT2022zod.}
\begin{tabular}{l c c}
\hline\hline
\textbf{Parameter} & \textbf{Prior Range} & \textbf{Best-fit Value} \\
\hline
$M_{\rm BH}\ (M_\odot)$  & $[10^{4},\,5\times10^{8}]$       & $(1.3^{+0.8}_{-0.5})\times10^{6}$ \\
$M_{*}\ (M_\odot)$       & $[0.01,\,30]$                    & $0.22^{+0.27}_{-0.11}$ \\
$\eta$                   & $[10^{-4},\,0.4]$                & $0.006 \pm 0.003$ \\
$b$                      & $[1.0,\,2.0]$                    & $1.27^{+0.18}_{-0.16}$ \\
\hline
\end{tabular}
\label{tab:mosfit_bestfit}
\end{table}

In \texttt{MOSFiT}, we employed the UltraNest sampler together with the default priors. The fit was performed over the interval MJD 59860–59905. The resulting posterior distributions indicate a maximum black hole mass of order $10^6$$M_\odot$, consistent with the disruption of a low-mass star ($M_* < 0.4$ $M_\odot$), see Table \ref{tab:mosfit_bestfit}. 

\begin{figure*}
    \centering
    \includegraphics[width=6in]{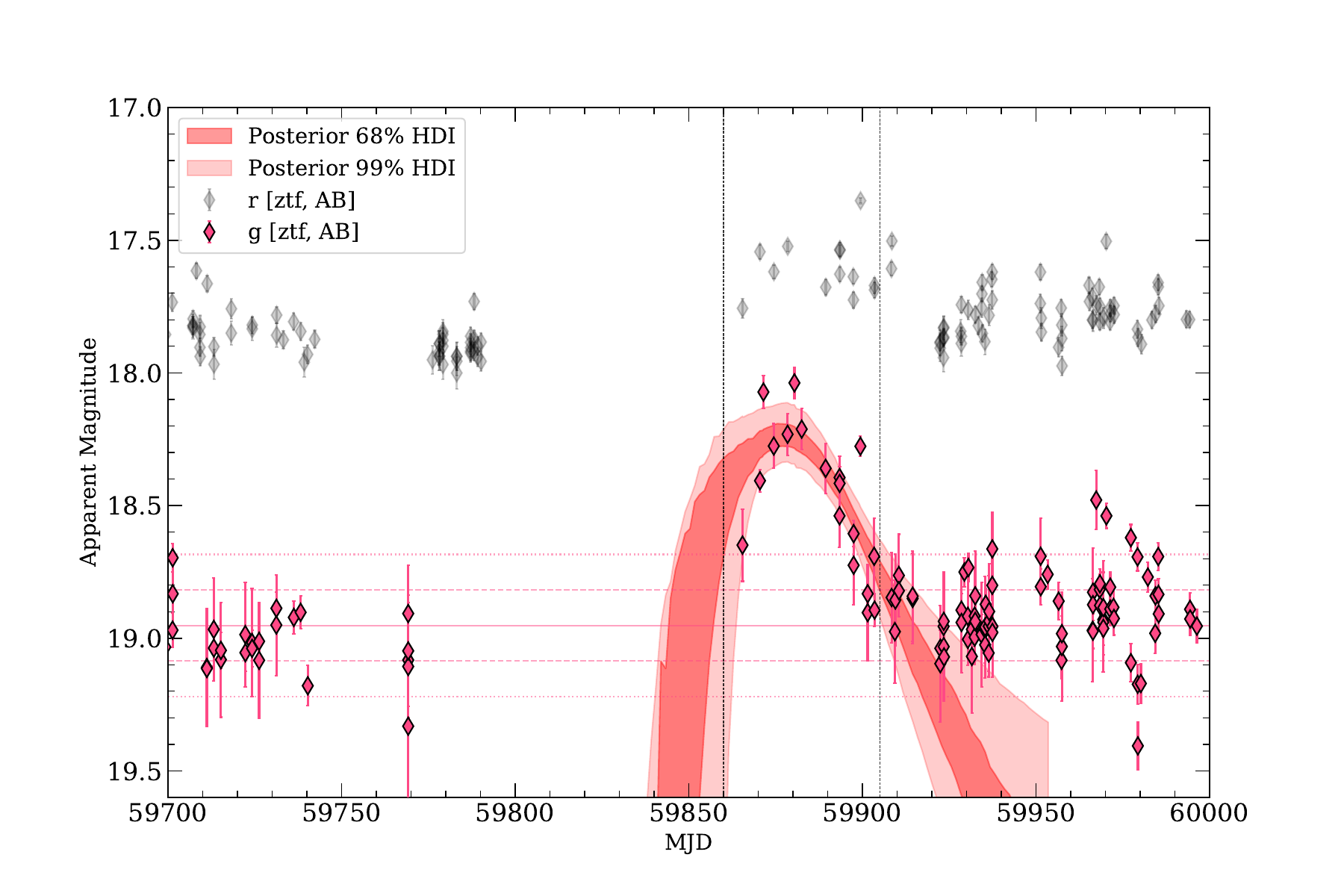}
    \caption{Fitted light curve of AT2022zod modeled with \texttt{MOSFiT}. Shaded regions denote the 68\% and 99\% highest-density intervals of the posterior-predictive samples. The dotted vertical lines indicate the time range over which the data were fitted.}
    \label{fig:mosfit}
\end{figure*}

While any mass obtained from modeling the lightcurves will be highly model dependent \citep{KScomp}, we emphasize that both fitting approaches consistently point to AT2022zod originating from a massive black hole in the intermediate-mass regime, with an inferred mass that is incompatible with the galaxy’s central SMBH.
 
\subsubsection{Ruling out non-TDE origins of AT2022zod}\label{ruling_out_nontde}

To further guide our characterization of this candidate, we employed the \texttt{Redback} software package \cite{REDBACKSarin2024}, which performs Bayesian inference to model electromagnetic transients using a wide range of physically and empirically motivated models. 
Prior to the analysis, the light curve was preprocessed by restricting the dataset to the temporal window spanning approximately 15~days before and 40~days after the peak brightness, ensuring that the analysis focused on the well-sampled phase of the transient evolution. 
This interval was chosen based on the observed characteristics of AT2022zod, which exhibits a rise time of $\sim13$~days. 
The first observation in our dataset occurs roughly 16~days before the inferred peak (MJD), and therefore provides a good approximation of the explosion epoch, lying within $1\sigma$ of the explosion time derived from the generic flare fit.

We used \texttt{Redback} to perform a Bayesian model comparison across several physically motivated scenarios of increasing complexity. 
A full discussion of each model and its underlying assumptions lies beyond the scope of this paper; however, for completeness we summarize in Table \ref{tab:model_evidence} the main energy sources represented by each model and their typical associated transient.  For all models, we adopted the default prior distributions provided by \cite{REDBACKSarin2024} and \cite{Guillochon2018_mosfit}.

\subsubsection{AGN Flaring}\label{agn_flaring}
We searched the entire ZTF D23 light curve of AT2022zod for previous or posterior brightening and outbursts similar to this transient. Photometry is available in $r-$ and $g-$bands between MJD 28202.28133 (2018-03-25) and 60608.47095 (2024-10-25). 
To identify flux variations such as flares or outbursts in the light curve, we applied a sigma-clipping–based anomaly detection.The light curve was first sorted by time and smoothed using a median filter (window size of 7 epochs) to estimate the quiescent baseline. Residuals were then computed as the difference between the observed magnitudes and the smoothed baseline. We used the median absolute deviation (MAD) to estimate a robust scatter $\sigma$, and defined significant brightening events as epochs with residuals larger than $3\sigma$ relative to the baseline and with a duration of $> 2$ days (following the approximate cadence of the survey) after grouping single flares.

The only significant event retrieved from this analysis was AT2022zod itself. We can therefore rule out stochastic variability or outbursts  such as the ones seen in AGNs \cite[e.g., ][]{1997ARA&A..35..445U}, at least for the time period analyzed. 

\subsubsection{Periodic Tidal Disruption Events}

A periodic tidal disruption event (pTDE; \cite{pTDE10.1093/mnras/stab802}) could, in principle, reproduce the observed timescales and energetics of AT2022zod. However, proposed pTDEs are generally identified through recurrent flaring activity \cite[e.g.,][]{pTDESun2025,pTDEJi2025}. In this case, we detect only one outburst, and archival data show no evidence for an earlier, longer-duration flare over the preceding five years. This lack of recurrence makes a pTDE interpretation unlikely, thus we also disfavor this scenario.

\subsubsection{Compact binary mergers}
The observed properties of AT2022zod rule out a neutron star–black hole (NS–BH) or binary neutron star (BNS) merger as the progenitor system. Given its duration, brightness and distance, as well as the absence of multi-wavelength counterparts, a compact merger progenitor is highly unlikely. The radioactive decay of unstable nuclei synthesized in the neutron-rich ejecta of a compact binary merger powers a rapidly evolving, approximately isotropic thermal transient known as a kilonova. An ultraviolet-optical-infrared transient, whose brightness peaks $\simeq$ 2 to 3 days after the merger \citep{Li1998,Darc2024}. However, in the case of AT2022zod, the optical flare persisted for approximately 30 days, significantly longer than the typical $\leq$ 10-day timescale observed in confirmed kilonovae such as AT2017gfo \citep{Metzger2019}.

Another interesting hypothesis to evaluate is whether this event originated from a binary black hole (BBH) merger embedded within an AGN disk. Such transients, are thought to arise from a highly spinning black hole remnant that receives a gravitational recoil kick after merger  (\cite{Graham2020, Darc2025, RodrguezRamrez2025}). The recoiling remnant travels into a denser region of the AGN disk, where it can efficiently accrete gas and potentially launch relativistic jets. Under favorable conditions, the interaction of these jets with the surrounding disk material may produce a detectable transient that can temporarily dominate over the intrinsic AGN emission. This emission can be easily mistaken by AGN variability in the optical band. To rule out this scenario, we cross-matched the source position with the Milliquas catalog \citep{Flesch_2023} within a 4'' radius and also with the DESI catalog, both of which returned a null probability for association with an AGN. Furthermore, theoretical models predict that BBH mergers within AGN disks typically occur around supermassive black holes with masses in the range $10^{7}$–$10^{8},M_\odot$, and that most configurations produce electromagnetic counterparts peaking within $\sim 50$ days of the gravitational-wave detection (\cite{Darc2025}). In our case, the lack of AGN lines in the optical spectrum disfavors an AGN-disk environment, Therefore, we consider this scenario unlikely to explain the observed candidate.

\subsubsection{Supernovae}
Another potential source of contamination in our analysis arises from supernovae (SNe), which occur far more frequently than TDEs and can, under certain circumstances, exhibit optical light curves that mimic the fast, luminous flares expected from accretion-powered transients such as TDEs. The advent of wide-field, high-cadence surveys such as ZTF, Pan-STARRS \citep{PanSTARRSKaiser2010}, and ATLAS \citep{ATLASTonry2018}, has enabled the systematic characterization of SN populations as a function of their rise times (the interval between explosion and peak luminosity), and their peak absolute magnitudes. These physical parameters provide a valuable framework for distinguishing SNe from other fast optical transients solely based on their photometric evolution.

Among the diverse SN subclasses identified in the literature, only a small fraction display rapid rises ($t_{rise}<15$~days) without a plateau phase or secondary peak. Superluminous supernovae (SLSNe) represent a population of SNe whose peak luminosities are significantly higher than those of canonical SNe, typically spanning the absolute magnitude range of $-19.8 \lesssim M_{\mathrm{abs}} \lesssim -23$~mag (\cite{GalYam2019}). The physical mechanisms responsible for powering SLSNe are not yet fully understood, though four primary energy sources have been proposed: (1) radioactive decay of $^{56}$Ni, (2) interaction between the ejecta and dense circumstellar material (CSM), (3) rapidly rotating magnetar spin-down, and (4) fallback accretion onto a compact remnant \citep{Moriya2018}. The mean rise time of hydrogen-poor (Type~I) SLSNe is approximately $41 \pm 18$~days in the $g$ band \citep{Chen2023}, with typical peak absolute magnitudes reaching $M_{g} \approx -21$~mag.

Notably, SN2018bgv, a fast-evolving hydrogen-poor SLSN observed by the ZTF, exhibited a rapid rise of $\sim 10$ days and reached a peak magnitude of $M_{g} \approx -21$ mag \citep{Lunnan2020}. Similarly, iPTF16asu, classified as a TypeIc SLSN (SLSN-Ic; \cite{iPTF16asuWang2022}), was an exceptionally luminous and rapidly evolving event, achieving $M_{g} = -20.4$~mag after a rise of only $\sim 4$ days \citep{Whitesides2017}. Lastly, SN~2021lwz represents another peculiar luminous Type Ic-like SNe, reaching a peak magnitude of $M_{g} \sim -20.1$~mag (AB system) within $\sim 7$days of explosion.
Most of these transients were interpreted as being powered by a hybrid mechanism combining magnetar spin-down with radioactive heating from $^{56}$Ni decay \citep{iPTF16asuWang2022, Whitesides2017}.

By comparison, AT~2022zod exhibits a faster and more luminous evolution than most types of SNe, with a rest-frame rise time of $\approx 13$~days and a peak absolute magnitude of $M_{g} \approx -20.5$~mag, corresponding to a rise rate of $\approx 0.66$~mag~day$^{-1}$. Such rapid evolution and high luminosity place it at the extreme end of the supernova population and suggest that alternative mechanisms: such as magnetar spin-down, may be responsible for its observed properties, if it is a supernova. However, as we demonstrate, the lightcurve can be well-fitted by a fallback accretion model such as a TDE.

\begin{figure}
    \centering
    \includegraphics[width=1.0\linewidth]{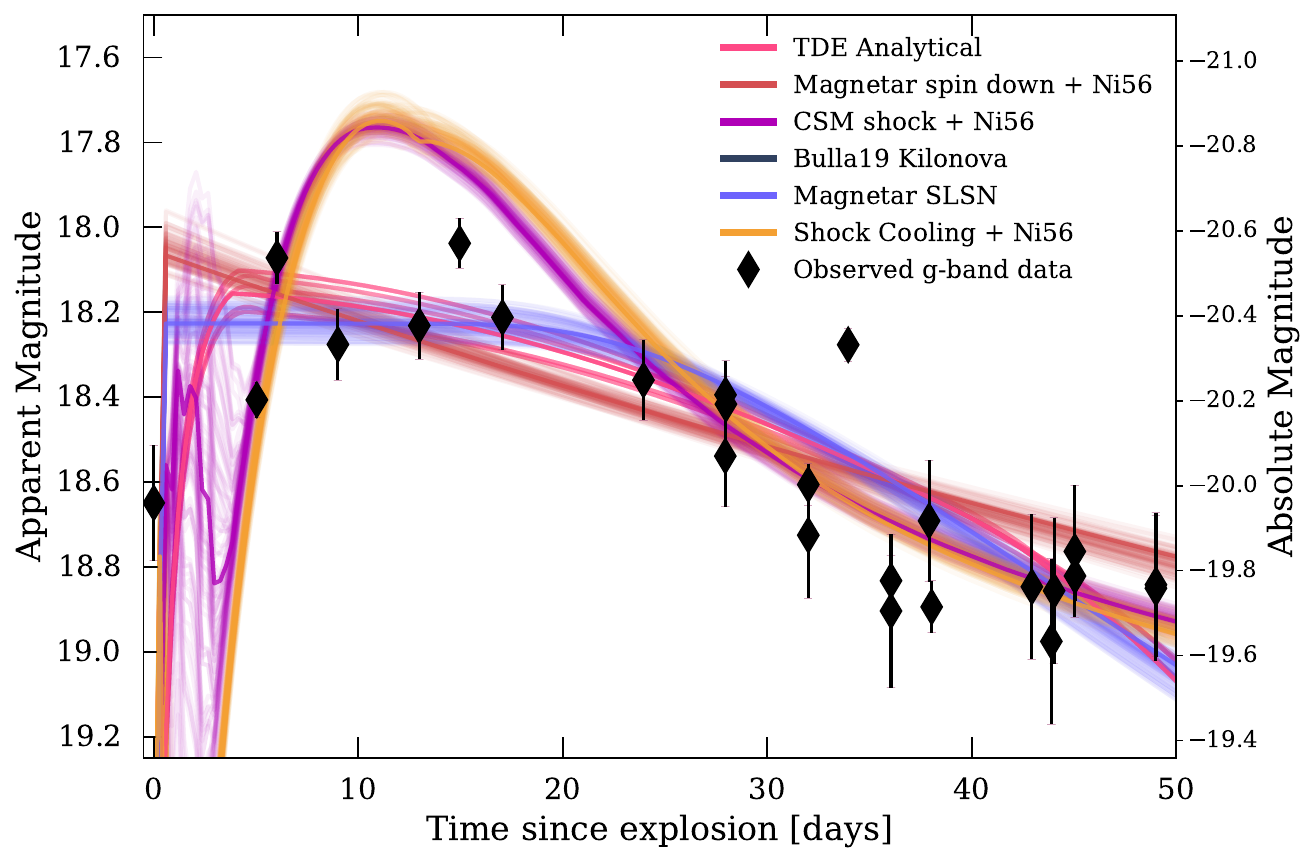}
    \caption{Light-curve modeling of AT2022zod performed with the Redback framework. For each emission scenario, 100 posterior samples are shown in distinct colours, with the maximum-likelihood model highlighted in a solid, high-opacity curve. The observed photometry is displayed in black. The Bulla19 kilonova model is omitted because its predicted apparent brightness at the source distance falls far below the data. None of the explored models provides an adequate description of the observed light curve.}
    \label{fig:redback}
\end{figure}

We present the \texttt{Redback} light-curve fits in Fig. \ref{fig:redback} and the corresponding model evidences in Table \ref{tab:model_evidence}. None of the tested configurations provide an adequate description of AT2022zod. Our analysis indicates that AT2022zod does not exhibit signatures of AGN variability or a supernova, making a TDE from a MBH one of the most viable explanations for the flaring source. 

\begin{table*}
\centering

\begin{tabular}{lccc}
\hline
\hline
\textbf{Model Name} & \textbf{log$_{10}(Z)$} & \textbf{Description} & \textbf{Applicable Types} \\
\hline

\texttt{shock\_cooling\_and\_arnett} & $-66.78 \pm 0.10$ & Shock cooling + NiCo decay & IIb, Ib/c \\[3pt]
\texttt{csm\_shock\_and\_arnett} & $-102.47 \pm 0.09$ & CSM shock breakout + NiCo decay & SLSN-II, IIn, Icn \\[3pt]
\texttt{magnetar\_nickel} & $-68.60 \pm 0.08$ & Magnetar + NiCo decay & SLSN-I, luminous Ic \\[3pt]
\texttt{general\_magnetar\_slsn} & $-63.22 \pm 0.09$ & General magnetar engine + diffusion & SLSN-I \\[3pt]
\texttt{bulla\_bns\_kilonova} &  $-207124.5 \pm 1.7$ & Multi-component BNS kilonova & Kilonova (BNS) \\[3pt]
\texttt{tde\_analytical} &  $-86.78 \pm 1.02$  & Analytical TDE fallback  ($t^{-5/3}$) & TDE \\[3pt]

\hline
\end{tabular}
\caption{Bayesian evidence comparison of transient models fitted with \texttt{Redback}. Listed are the logarithmic evidences ($\log_{10} Z$), brief model descriptions, and typical applicable transient types.}
\label{tab:model_evidence}
\end{table*}

\section{Results}

We identified the event AT2022zod, located in an elliptical galaxy with spectroscopic redshift from SDSS z=0.11. Based on a Bayesian hierarchical model of the M-sigma relation implies that the host likely contains a supermassive black hole of mass
$M_\bullet \simeq 1.0 \times 10^{8}\,M_\odot$ SMBH.

The event lasted roughly 30 days, with a rise time of 13.24 $\pm 3.68$ days. Although no multiwavelength follow-up was obtained, we modeled the optical light curve and used the inferred flare timescales to rule out several classes of fast extragalactic transients, including kilonovae, compact-binary mergers, stellar-mass black hole TDEs, and periodic TDEs. 
An AGN flare would be the most obvious explanation, however, we disfavor this scenario as this transient event is the only flaring activity in the full ZTF lightcurve.

Our lightcurve modeling with TiDE suggests that AT2022zod may be consistent with the upper mass range of an IMBH. However, we caution that this is assuming that IMBH TDEs behave like a scaled down version of SMBH TDEs. We note the discussion from \citep{Chang2025}, that it is possible for IMBH TDEs to happen on both longer \citep{dai, he21} and shorter \citep{margutti19} timescales than SMBH TDEs.  The model fits implied by TiDE suggests a black hole mass around $10^5 M_\odot$. The lightcurve fitting and comparison with observed TDEs suggests that duration of the observed flare of AT2022zod is inconsistent with TDE by a $10^8 M_\odot$ SMBH, at least based on our current understanding of TDE lightcurve models.

We propose the following interpretation: the 30-day optical flare of AT2022zod is too long for a kilonova, too luminous to resemble a typical supernova, and inconsistent with a TDE powered by the galaxy’s central SMBH. A periodic TDE from the SMBH is also disfavored: no recurrent flares are present in the five years of ZTF monitoring prior to AT2022zod, nor in the two years following it. Although the recurrence timescales of periodic TDEs remain uncertain, the small number of confirmed cases suggests intervals of only a few years, making the absence of additional flares noteworthy. Similarly, the absence of any other variability in the host galaxy further argues against normal AGN activity. While the light-curve modeling alone cannot tightly constrain the black-hole mass, and thus the evidence for an intermediate-mass black hole remains tentative, the overall behaviour of the flare is most naturally generated by disruption from a MBH.

We further speculate on the possible origins of the flare. Given the nominal ZTF astrometric uncertainty, the transient could plausibly have occurred anywhere up to around 3 kpc from the galaxy center. If AT2022zod originated within the host nucleus, while MBHs may be captured by the AGN disk \cite[e.g.,][]{2023MNRAS.522.5393N}, the host galaxy's observed properties disfavor an AGN-disk environment. However, it is possible for IMBHs up to $10^4 M_\odot$ to reside within the central nuclear star cluster, and form via collisions within the nuclear star cluster  \cite[e.g.,][]{Rose2022}. If so, this has the potential to form an intermediate mass ratio inspiral. Another alternative explanation if AT2022zod is truly an IMBH is that it may have originated in a globular cluster as suggested by several decades of simulations \citep{2002ApJ...576..899P,2025arXiv251100200P}, and we would expect AT2022zod's host environment to be home to an extensive globular cluster system. Nevertheless, we argue that the most likely origin is a UCD embedded in the host galaxy, hosting its own massive black hole. This environment closely resembles those inferred for several other TDEs associated with MBHs \citep{Lin18, Jin25, Angus22, 2025arXiv251012572P}. Such an off-nuclear location is also consistent with the predictions by \cite{Bellovary10} and with the distribution of UCDs observed in the Local Volume \citep{2014Natur.513..398S, 2025ApJ...991L..24T, 2025arXiv251109641S}.

  \section{Conclusions}
  
AT2022zod is an unusually luminous transient at  z=0.11, with an observed duration of roughly 30 days, reaching a much higher peak luminosity than most events with that timescale. From the photometric properties, and through fits with standard light-curve models, we strongly disfavor AGN variability, supernovae, kilonovae, and periodic TDE scenarios. Instead, our modelling  suggests that AT2022zod may be either a tidal disruption event by a massive black hole in the intermediate-mass range or by a star disrupted by the central SMBH on a non-parabolic orbit. Given the observed timescales and the results of our \texttt{TiDE} light-curve analysis, we find it more plausible that AT2022zod originates from a second, less massive black hole within the system.

If the transient is nuclear, i.e., originates in the galaxy center, the central engine may be a MBH co-residing in the host’s nuclear star cluster with the central SMBH. Alternatively, If the black hole's true mass is less than what the light curve models indicate, i.e, an IMBH, AT2022zod's black hole may have formed in a globular cluster \cite{2025arXiv251100200P}. However, the most natural explanation is that AT2022zod originates from an ultra-compact dwarf acquired by the host galaxy early on in the Universe. Such systems may offer valuable insight into the behavior and evolutionary role of UCDs at higher redshifts.

While the rates of IMBH TDEs remain poorly constrained, the Rubin Observatory is expected to detect such events in significant numbers. TDE Events like AT2022zod,  unclassified by current alert brokers, will be essential for improving the MBH census up to higher redshifts. We note that while AT2022zod's observed properties are consistent with a TDE by an off-nuclear MBH, it was not discovered in time to obtain multiwavelength followup to confirm the full nature of the emission mechanism. Hence, rapid identification of similar events in the Vera Rubin era promises to provide fascinating information about these new puzzling events.

In the Rubin era, building a larger statistical sample of short-duration TDEs will be crucial for identifying a sufficiently robust population of IMBHs to distinguish their formation pathways and evolutionary histories, thereby advancing our understanding of SMBH  formation in the early Universe. Progress will require both theoretical advances to better model IMBH-driven TDEs, enabling tighter mass constraints from light curves, and observational strategies that extend TDE searches to the full radial extent of elliptical galaxies, where IMBHs may be more likely to reside towards the outskirts.  

Finally, we note that, despite its unusual properties, AT2022zod remained effectively unnoticed in ZTF data for more than two years and was discovered only through a systematic search for flares in elliptical galaxies with no prior constraints on position within the host, minimum brightness, or light-curve morphology.  This discovery highlights that future efforts to constrain black hole populations and their evolution should remain agnostic with respect to the properties of TDEs. As emphasized by \cite{2025arXiv251119016Q},  overly restrictive selection strategies risk a substantial loss of opportunity, particularly of more peculiar events. Given the data volume and complexity expected from Vera Rubin observatory, searches will necessarily be targeted to explicit scientific goals, but they must also retain sufficient flexibility to accommodate unexpected phenomena; otherwise, we may fail to optimally exploit data sets we have invested so heavily to obtain.

\section*{Acknowledgements}

The authors thank R. Hirai and I. Mandel for helpful discussion. ACS acknowledges support from FAPERGS (grants 23/2551-0001832-2 and 24/2551-0001548-5), CNPq (grants 314301/2021-6, 312940/2025-4, 445231/2024-6, and 404233/2024-4), and CAPES (grant 88887.004427/2024-00). SS acknowledges support from the UK Science and Technology Facilities Council (STFC) via the grant ST/X508408/1. LN acknowledges FAPESP grant number 2024/07281-0. This work is a result of the COIN Residence Program \#8\footnote{\url{https://cosmostatistics-initiative.org/residence-programs/crp-8-08-15-sep-2025-brazil/}}, held in Armação dos Búzios, Brazil, from 8 to 15 September 2025 and supported by the Brazilian agency Conselho Nacional de Desenvolvimento Científico e Tecnológico (CNPq). The Cosmostatistics Initiative\footnote{\url{https://cosmostatistics-initiative.org/}} (COIN) is an international network of researchers whose goal is to foster interdisciplinarity inspired by astronomy. This work made use of the Fink resources. Fink is supported by LSST-France and CNRS/IN2P3.
%

\vspace{5mm}






\bibliography{sample631}{}
\bibliographystyle{aasjournal}



\end{document}